\documentclass[aps,prc,twocolumn,showpacs,nofootinbib]{revtex4}

\usepackage{graphics}
\usepackage{color}
\usepackage{amssymb}
\usepackage{amsmath}
\usepackage{pifont}

\newcommand{\sss}{\scriptscriptstyle}

\begin{document}

\title{Constraints on background contributions\\
from $K^+ \Lambda$ electroproduction}
\author{S. Janssen}
\email[]{stijn.janssen@rug.ac.be} 
\author{J. Ryckebusch}
\author{T. Van Cauteren}

\affiliation{Department of Subatomic and Radiation Physics, Ghent
University, Belgium}

\date{\today}

\begin{abstract}
Results for response functions for kaon electroproduction on the
proton are presented.  A tree-level hadrodynamical model is adopted
and it is shown that some of the electroproduction response functions
are particularly powerful with the eye on gaining control over the
parameterization of the background diagrams. The existing data set for
the $p(e,e'K^+)\Lambda$ reaction appears to rule out the use of a $g_{K^+
\Lambda p}$ coupling constant beyond the boundaries of softly broken
SU(3) flavor symmetry.  Also the use of soft hadronic form factors,
which has been proposed as a valid alternative for a hadrodynamical
description of the $p(\gamma,K^+)\Lambda$ data in the resonance
region, seems to be disfavored by the magnitude of the measured
$p(e,e'K^+)\Lambda$ cross sections.
\end{abstract}


\maketitle

In studies of the baryon resonance spectrum, the electromagnetic
production of mesons is a privileged reaction.  Traditionally, most of
the efforts have been directed towards the pion production channels.
Sparked by major experimental efforts at accelerator facilities like
Jlab, ELSA, SPring-8 and GRAAL there is growing interest in other
meson production reactions like $\eta N$, $\omega N$, $K \Lambda$ and
$K \Sigma$. Amongst them, the strangeness production channels
constitute a special class of reactions. Indeed, the involvement of
the strange $s \overline{s}$ quark anti-quark pair in the reaction
dynamics opens an additional window to study nucleon resonances. The
SAPHIR collaboration at ELSA \cite{Tran} has measured
$p(\gamma,K^+)\Lambda$ and $p(\gamma,K^+)\Sigma^0$ differential cross
sections and recoil polarizations from threshold up to photon energies
of 2 GeV. At present, the published $p(e,e' K^+)\Lambda$ data set is
rather sparse with a few results from measurements in the seventies at
Orsay \cite{Brown_cea}, Cornell \cite{Bebek_74,Bebek_77} and DESY
\cite{Azemoon} and recent data from Hall C at Jlab
\cite{Niculescu,Mohring}. In the near future, however, concerted
efforts at the Jlab facility will greatly improve on this situation.

In Ref.\cite{Janssen_backgr}, we have shown that an important fraction
of the $p + \gamma \rightarrow K^+ + \Lambda$ reaction dynamics in the
resonance region stems from background contributions. In the same
work, we have discussed results obtained with three different schemes
to deal with the background Feynman diagrams.  We concluded that with
the existing amount of $p(\gamma,K^+)\Lambda$ data, one is not able to
put one of these schemes forward as most adequate.  The extracted
resonance information, however, turns out to be rather sensitive to
the model choices with respect to the parameterizations of these
background diagrams. In this work, our hadrodynamical model for
$p(\gamma,K^+)\Lambda$ photoproduction will be applied to the
corresponding electroproduction process.  The cross section for the
virtual photon induced reaction can be decomposed as:
\[
\frac{ d \sigma}{ d \Omega} = \frac{d \sigma_{\sss T}}{ d \Omega} +
\epsilon \frac{d \sigma_{\sss L}}{ d \Omega} + \epsilon \frac{d
\sigma_{\sss TT}}{ d \Omega} \cos 2 \phi + \sqrt{\epsilon
\left(\epsilon + 1 \right)} \frac{d \sigma_{\sss TL}}{ d \Omega} \cos
\phi
\]
We wish to demonstrate that some of the four $p(e,e'K^+)\Lambda$
response functions offer good prospects to constrain the ambiguities
in the description of the background which emerge from analyzing the
real-photon data.

The different ingredients in the reaction dynamics implemented in our
$p(e,e'K^+)\Lambda$ calculations are essentially identical to the ones
adopted for the description of the $p(\gamma,K^+)\Lambda$ process
reported in Ref.~\cite{Janssen_backgr}.  This implies that we start
from a given set of interaction Lagrangians with each term having its
characteristic coupling constant. From there, we derive both the
longitudinal and transverse electromagnetic amplitudes. We wish to
stress that also the resonances are described in the Lagrangian
formalism and that no multipole decomposition gets introduced as is
commonly done in calculations for $\pi$ and $\eta$
electroproduction. The tree-level Feynman diagrams implemented in the
calculations include the usual Born terms and the $K^*(892)$
and $K_1(1270)$ mesons in the $t$-channel. As will be pointed out
below, at some point two $\Lambda^*$ resonances ($S_{01}(1800)$ and
$P_{01}(1810)$) will be introduced in the $u$-channel. All those terms
constitute the so-called background. In the $s$-channel, the nucleon
resonances $S_{11}(1650)$, $P_{11}(1710)$, $P_{13}(1720)$ and
$D_{13}(1895)$ are retained. Note that the $D_{13}(1895)$ resonance is
not listed in the Particle Data Group booklet \cite{PDG} but is a
candidate for a ``new'' resonance.  A substantial improvement in the
quality of the description of the $p(\gamma,K^+)\Lambda$ data was
reached after including this resonance \cite{Mart2,Janssen_role_hyp}.

In electroproduction processes, an additional form factor gets
introduced at the electromagnetic vertices. For the Pauli and Dirac
form factors of the proton, the parameterization of Lomon \cite{Lomon}
is adopted. For the (transition) form factors of the $\Lambda$, $K$
and the $N^*$, $K^*$, and $\Lambda^*$ resonances, no well established
parameterizations are currently available.  Therefore, we rely on the
predictions of a relativistic constituent-quark model calculation by
the Bonn group \cite{Munz,Merten} for the $\Lambda$, $K^+$ and $K^*$
form factors. For the $N^*$ and $\Lambda^*$ transition form factors,
we use a dipole form with one universal cutoff mass of 0.84 GeV. For
the $K_1$, a monopole form with cutoff mass of 0.6 GeV was used.
These electromagnetic cutoff masses are the only extra numbers
entering our electroproduction calculations. All other parameters are
fixed by constraining the model against the SAPHIR data at the real
photon point. The sensitivity of the observables to the values of the
cutoff masses in the electromagnetic form factors will be discussed
below. In order to preserve gauge invariance at the level of the Born
terms after introducing electromagnetic form factors, the gauge
restoration procedure of Gross and Riska is adopted
\cite{Gross}. Results with alternative schemes will be discussed
below.

\begin{figure}
\resizebox{0.45\textwidth}{!}{\includegraphics{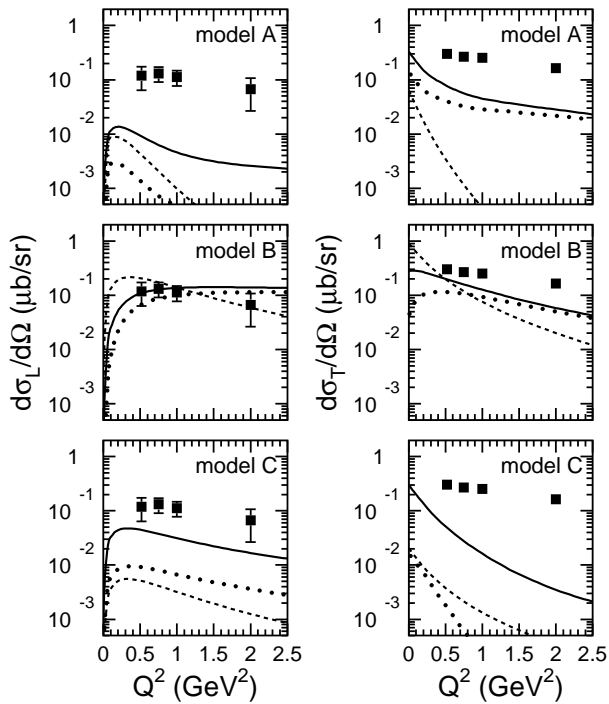}}
\caption{\label{fig:t_l_cont}Model predictions for the $Q^2$
dependence of the longitudinal and transverse $p(e,e'K^+)\Lambda$
response functions at $W$ = 1.84 GeV and $\cos \theta$ = 1. The
three different panels are obtained with the background schemes of
model A, B and C. The dashed curve represents the contribution from
the Born terms, the dotted curve the entire background and the solid
curve the sum of the complete background and resonance diagrams. The
data are from Ref.~\protect\cite{Mohring}.}
\end{figure}

The magnitude of the Born contributions to the computed
$p(\gamma,K^+)\Lambda$ strength is essentially determined by an
effective coupling of the type \cite{Feuster2}:
\begin{equation}
G_{K^+ \Lambda p} \equiv g_{K^+ \Lambda p} \cdot F_h \left(x,
\Lambda_h \right) \;,
\label{eq:general_cc}
\end{equation}
where $g_{K^+ \Lambda p}$ can be related to the pion strength $g_{\pi
N N}$ through SU(3) flavor symmetry. Further, $F_h \left(x, \Lambda_h
\right)$ denotes the hadronic form factor, $\Lambda_h$ the cutoff
parameter and $x \equiv \left( s, t, u \right)$ is the Mandelstam
variable at the hadronic vertex. We use a dipole parameterization for
$F_h \left(x, \Lambda_h \right)$ \cite{Haberzettl_gauge,Davidson}.
Hard cutoff masses (typically, $\Lambda_h \geq$ 1.5 GeV) correspond
with $F_h(x, \Lambda_h) \approx$~1 over the entire resonance region.
The background terms on their own overpredict the
$p(\gamma,K^+)\Lambda$ data dramatically when the effective coupling
$G_{K^+ \Lambda p}$ goes out from a modest SU(3) flavor symmetry
breaking, at the same time keeping the impact of the hadronic form
factor temperate by fixing $\Lambda_h \geq$ 1.5 GeV. This situation
can be rectified through decreasing the coupling constant $g_{K^+
\Lambda p}$ by several factors, thereby putting forward strong SU(3)
flavor symmetry breaking \cite{Hsiao}. Alternatively, the hadronic
form factor can be adjusted in such a manner so as to sufficiently
reduce $G_{K^+ \Lambda p}$. In practice, this amounts to adopting
smaller values of $\Lambda_h$, thereby amplifying the dependence of
the results on the hadronic form factors \cite{Haberzettl}.  In
practice, acceptable levels of the computed Born strength, which we
define as being of the same order of magnitude as the measured cross
sections, require cutoff masses $\Lambda_h$ which approach the kaon
mass, leading to a very unsatisfactory situation from the
field-theoretic point of view. Both above mentioned manipulations
amount to effectively reducing $G_{K^+ \Lambda p}$, either through
adjusting the coupling constant or the hadronic form factor or a
combination of both, thereby making assumptions which are rather
questionable. Therefore, instead of adjusting the effective coupling
$G_{K^+ \Lambda p}$, we have suggested an alternative procedure
consisting of introducing hyperon resonances as a more natural
mechanism to counterbalance the Born strength
\cite{Janssen_role_hyp}. Those $u$-channel diagrams are observed to
interfere destructively with the other background terms. In this way,
a qualitatively good description of the $p(\gamma,K^+)\Lambda$ data
can be reached, without the need of introducing rather questionable
values for the $g_{K^+ \Lambda p}$ coupling and/or $\Lambda_h$.

\begin{figure}
\resizebox{0.45\textwidth}{!}{\includegraphics{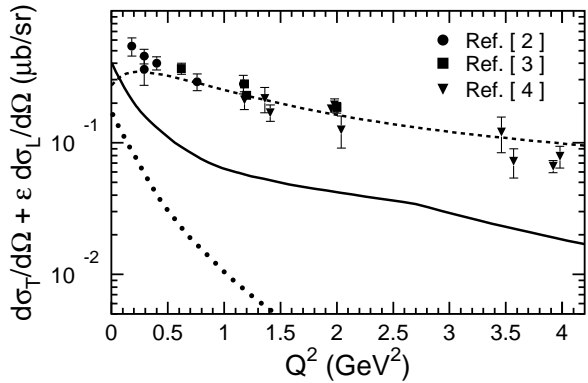}}
\caption{Calculations for the $Q^2$ dependence of the $\phi$-averaged
  $p(e,e'K^+)\Lambda$ response functions at $\langle W \rangle$ = 2.15
  GeV and forward $\theta$ angles. The solid, dashed and dotted line
  are from model A, B and C, respectively. Data is from
  Refs.~\protect\cite{Brown_cea,Bebek_74,Bebek_77}.\label{fig:cs_t+l}}
\end{figure}

The three aforementioned ways of treating the background diagrams are
labeled as model A, B and C.  Model A adopts soft hadronic form
factors with $\Lambda _h$ approaching the kaon mass. Model C uses
$g_{K^+ \Lambda p}/\sqrt{4 \pi} \approx -0.4 $, which is almost ten
times smaller than the prediction based on SU(3) flavor
symmetry. Whereas models A and C lower $G_{K^+ \Lambda p}$, model B
introduces hyperon resonances in the $u$-channel and attributes a
secondary role to the hadronic form factors, at the same time
respecting the constraints on $g_{K^+ \Lambda p}$ imposed by SU(3)
flavor symmetry. As pointed out in Ref.~\cite{Janssen_backgr}, all
three models lead to a similar quality of agreement between the
calculations and the $p(\gamma,K^+)\Lambda$ data and none of the three
schemes could be put forward as favorable. In the electroproduction
calculations, however, large differences emerge between the
predictions of the three different background models.  This is made
clear in Fig.~\ref{fig:t_l_cont} showing model predictions for the
$Q^2$ dependence of the longitudinal and transverse
$p(e,e'K^+)\Lambda$ response functions at a particular value for the
invariant mass $W$ and the kaon cm angle $\theta$.  The background
models A and C are discerned to severely underestimate the
longitudinal and the transverse response. Model B, on the other hand,
provides a prediction of the magnitude and $Q^2$ dependence of both
observables which is far superior to what is obtained with models A
and C. The large variations between the predictions of the background
models can be better understood by decomposing the response functions
in contributions from the Born terms, the total background and
$s$-channel resonances. Then, it becomes apparent that a necessary
condition for arriving at a reasonable prediction of the data is that
the combined background diagrams already lead to response functions
which are of the order of the measured strength. As such, the
$p(e,e'K^+)\Lambda$ observables appear to provide direct access to the
background contributions and may eventually allow us to gain further
control over the value of $G_{K^+ \Lambda p}$.  Similar trends are
observed in Fig.~\ref{fig:cs_t+l} where the model calculations are
compared with the available $d \sigma_T + \epsilon d \sigma_L$ data
for the $\phi$-averaged cross section at forward $\theta$.  On the
basis of the comparisons displayed in Figs.~\ref{fig:t_l_cont} and
\ref{fig:cs_t+l}, we are tempted to conclude that the physical
assumptions underlying models A and C, which are compatible with the
existing $p(\gamma,K^+)\Lambda$ data set, are not supported by the
$p(e,e'K^+)\Lambda$ data.  This suggests that calculations based on
the introduction of soft hadronic form factors and/or a $g_{K^+
\Lambda p}$ coupling constant strongly deviating from SU(3)
predictions are completely off when it comes to predicting the cross
sections for the corresponding electroinduced process.

\begin{figure}
\resizebox{0.45\textwidth}{!}{\includegraphics{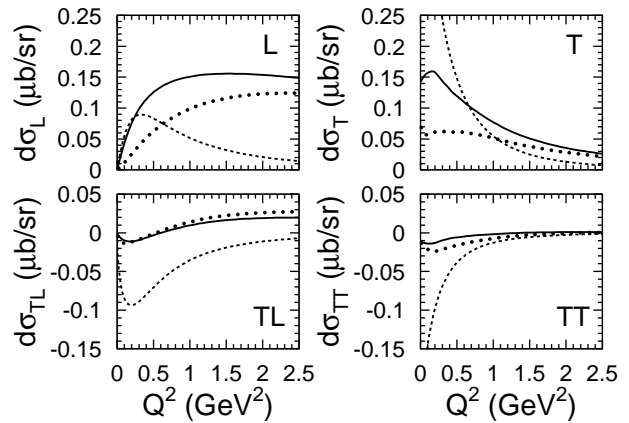}}
\caption{\label{fig:cs_th_av}Model calculations for the $Q^2$
  dependence of the $\theta$-averaged $p(e,e'K^+)\Lambda$ response
  functions at $W$ = 1.84 GeV. Background model B is adopted. Line
  conventions as in Fig.~\ref{fig:t_l_cont}.}
\end{figure}
It is worth stressing that the results contained in
Figs.~\ref{fig:t_l_cont} and \ref{fig:cs_t+l} refer to kinematics
whereby the kaon is emitted in a small cone about the direction of the
three-momentum transfer. In Fig.~\ref{fig:cs_th_av} we display the
corresponding $\theta$-averaged $d \sigma_L$, $d \sigma_T$, $d
\sigma_{TL}$ and $d \sigma_{TT}$ response functions. In the
angle-averaged responses, the strength directly related to the
$s$-channel resonances is at best of the same order as the one stemming
from the background diagrams and tends to decrease with increasing
$Q^2$. In that respect, the ratio of the background to resonance
strength in the $\theta$-averaged responses is rather similar to what
is observed at forward $\theta$ angles.
\begin{figure}
\resizebox{0.45\textwidth}{!}{\includegraphics{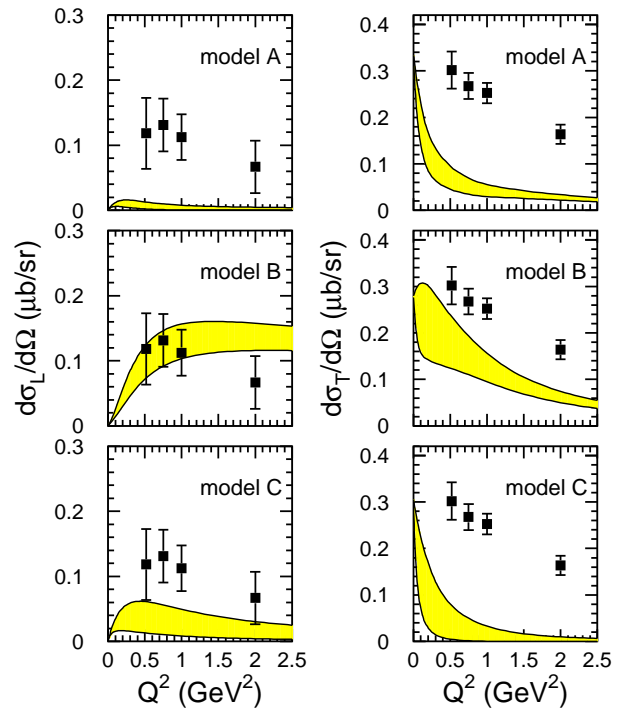}}
\caption{\label{fig:cs_t_l_emff} Sensitivity of the model calculations
of Fig.~\ref{fig:t_l_cont} to the $N^*$ electromagnetic form
factors. The shaded region displays the variations in the predictions
when using cutoff masses ranging between $0.4 \le \Lambda_{N^*} \le
1.0$~GeV.}
\end{figure}
\begin{figure}
\resizebox{0.45\textwidth}{!}{\includegraphics{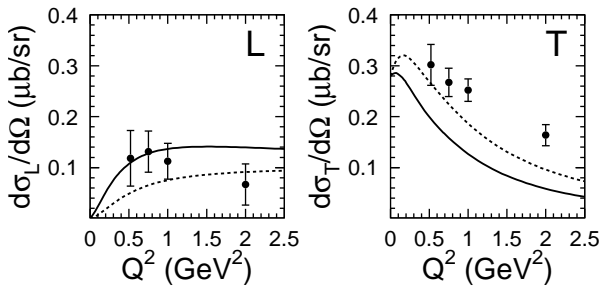}}
\caption{\label{fig:cs_t_l_gauge} Model calculations for the $Q^2$
  dependence of $d \sigma_L$ and $d \sigma_T$ with background model
  B. The solid line adopts the gauge restoration procedure of
  Ref.~\cite{Gross}, the dashed line uses modified form factors as
  explained in the text. Kinematics and data are as in
  Fig.~\ref{fig:t_l_cont}.}
\end{figure}

All results mentioned so far were obtained with a dipole
electromagnetic form factor for the $N^*$ resonances with a realistic
cutoff of 0.84 GeV. We now wish to investigate the sensitivity of our
results to this choice. Therefore, we have varied the $N^*$ dipole
cutoff mass between values which appear as upper and lower limits for
a physically realistic range. It seems that the results were rather
insensitive to those variations. This is shown in
Fig.~\ref{fig:cs_t_l_emff} where the shaded region indicates the
variation in the predictions when modifying the cutoffs in the range
$0.4 \le \Lambda_{N^*} \le 1.0$~GeV. All other electromagnetic form
factors in the dynamics of the background are kept fixed. From this
figure, we can conclude that reasonable changes in the functional
$Q^2$ dependence of the resonance couplings does not alter the marked
dominance of the background contributions.

Up to this point, all results are obtained with the gauge restoring
procedure of Ref.~\cite{Gross}. Within this scheme, one can use
different form factors for the proton and the kaon. Alternatively,
gauge invariance can be restored by using the same functional $Q^2$
dependence for the $F^p_1$ proton and $F_K$ kaon form factor. We have
investigated this option through averaging $F^1_p(Q^2)$ and $F_K(Q^2)$
and display some results in Fig.~\ref{fig:cs_t_l_gauge}. As was
already mentioned in e.g. Ref.~\cite{David}, the choices with respect
to the gauge restoring procedure and form factor parameterization have
a sizable impact on the results but are similar in size for the three
adopted background models.

We wish to stress again that the values of the coupling constants and
the hadronic cutoff parameters which enter our $p(e,e'K^+)\Lambda$
calculations are those which optimize the agreement between the
predictions and the data at $Q^2$ = 0. In order to exclude the
possibility that rather modest modifications in these
parameterizations of the coupling constants alter our findings, we
have refitted all the coupling constants of the three presented models
to a data set which includes both the photo- and electroproduction
data. The results of those fits barely deviate from the predictions
presented in Figs.~\ref{fig:t_l_cont} and \ref{fig:cs_t+l}.  Once
again we had to conclude that the background models A and C are
intrinsically incapable of reproducing the $p(e,e'K^+)\Lambda$ data.

Summarizing, we have extended our tree-level hadrodynamical analysis
of kaon photoproduction in the resonance region to $p(e,e'K^+)\Lambda$
processes.  Except for the $Q^2$ dependence of the electromagnetic
form factors, no new ingredients are introduced in the model.  In line
with our findings for the $p(\gamma,K^+)\Lambda$ reaction, also in the
corresponding electroproduction process a leading role is played by
the background diagrams. It was pointed out that a hadrodynamical
analysis of the $p(\gamma,K^+)\Lambda$ data at tree level faces
difficulties in pinning down those terms. As a matter of fact, we
propose that the longitudinal and transverse $p(e,e'K^+)\Lambda$
responses can serve as a reliable and powerful means of constraining
the parameters which enter the background diagrams. The recent Jlab
and older Cornell and Orsay $p(e,e'K^+)\Lambda$ data appear
incompatible with a hadrodynamical description based on a $g_{K^+
\Lambda p}$ coupling which is beyond the boundaries imposed by SU(3)
flavor symmetry. In addition, the use of soft hadronic form factors,
which after all provide an alternative for accounting for the
$p(\gamma,K^+)\Lambda$ data, leads to $p(e,e'K^+)\Lambda$ predictions
far below the level of the measurements. The introduction of hyperon
resonances in the $u$-channel, on the other hand, emerges as a valid
alternative for providing a consistent description of both
$p(e,e'K^+)\Lambda$ and $p(\gamma,K^+)\Lambda$ data, thereby
respecting the constraints imposed by SU(3) flavor symmetry. More data
on the separated response functions would help in further shedding
light on the issue of the background terms, and will eventually result
in reduced uncertainties in the extraction of the resonance parameters
from both the real and virtual photon kaon production data.


\begin{thebibliography}{21}
\expandafter\ifx\csname natexlab\endcsname\relax\def\natexlab#1{#1}\fi
\expandafter\ifx\csname bibnamefont\endcsname\relax
  \def\bibnamefont#1{#1}\fi
\expandafter\ifx\csname bibfnamefont\endcsname\relax
  \def\bibfnamefont#1{#1}\fi
\expandafter\ifx\csname citenamefont\endcsname\relax
  \def\citenamefont#1{#1}\fi
\expandafter\ifx\csname url\endcsname\relax
  \def\url#1{\texttt{#1}}\fi
\expandafter\ifx\csname urlprefix\endcsname\relax\def\urlprefix{URL }\fi
\providecommand{\bibinfo}[2]{#2}
\providecommand{\eprint}[2][]{\url{#2}}

\bibitem[{\citenamefont{{M.Q.~Tran {\em et al.}}}(1998)}]{Tran}
\bibinfo{author}{\bibnamefont{{M.Q.~Tran {\em et al.}}}},
  \bibinfo{journal}{Phys.~Lett.~B} \textbf{\bibinfo{volume}{445}},
  \bibinfo{pages}{20} (\bibinfo{year}{1998}).

\bibitem[{\citenamefont{{C.N.~Brown {\em et al.}}}(1972)}]{Brown_cea}
\bibinfo{author}{\bibnamefont{{C.N.~Brown {\em et al.}}}},
  \bibinfo{journal}{Phys.~Rev.~Lett.} \textbf{\bibinfo{volume}{28}},
  \bibinfo{pages}{1086} (\bibinfo{year}{1972}).

\bibitem[{\citenamefont{{C.J.~Bebek {\em et al.}}}(1974)}]{Bebek_74}
\bibinfo{author}{\bibnamefont{{C.J.~Bebek {\em et al.}}}},
  \bibinfo{journal}{Phys.~Rev.~Lett.} \textbf{\bibinfo{volume}{32}},
  \bibinfo{pages}{21} (\bibinfo{year}{1974}).

\bibitem[{\citenamefont{{C.J.~Bebek {\em et al.}}}(1977)}]{Bebek_77}
\bibinfo{author}{\bibnamefont{{C.J.~Bebek {\em et al.}}}},
  \bibinfo{journal}{Phys.~Rev.~D} \textbf{\bibinfo{volume}{15}},
  \bibinfo{pages}{594} (\bibinfo{year}{1977}).

\bibitem[{\citenamefont{{T.~Azemoon {\em et al.}}}(1975)}]{Azemoon}
\bibinfo{author}{\bibnamefont{{T.~Azemoon {\em et al.}}}},
  \bibinfo{journal}{Nucl. Phys. B} \textbf{\bibinfo{volume}{95}},
  \bibinfo{pages}{77} (\bibinfo{year}{1975}).

\bibitem[{\citenamefont{{G.~Niculescu {\em et al.}}}(1998)}]{Niculescu}
\bibinfo{author}{\bibnamefont{{G.~Niculescu {\em et al.}}}},
  \bibinfo{journal}{Phys.~Rev.~Lett.} \textbf{\bibinfo{volume}{81}},
  \bibinfo{pages}{1805} (\bibinfo{year}{1998}).

\bibitem[{\citenamefont{{R.M.~Mohring {\em et al.}}}(2002)}]{Mohring}
\bibinfo{author}{\bibnamefont{{R.M.~Mohring {\em et al.}}}},
  \bibinfo{journal}{nucl-ex/0211005}  (\bibinfo{year}{2002}).

\bibitem[{\citenamefont{Janssen et~al.}(2002)\citenamefont{Janssen, Ryckebusch,
  Debruyne, and {Van Cauteren}}}]{Janssen_backgr}
\bibinfo{author}{\bibfnamefont{S.}~\bibnamefont{Janssen}},
  \bibinfo{author}{\bibfnamefont{J.}~\bibnamefont{Ryckebusch}},
  \bibinfo{author}{\bibfnamefont{D.}~\bibnamefont{Debruyne}}, \bibnamefont{and}
  \bibinfo{author}{\bibfnamefont{T.}~\bibnamefont{{Van Cauteren}}},
  \bibinfo{journal}{Phys.~Rev.~C} \textbf{\bibinfo{volume}{65}},
  \bibinfo{pages}{015201} (\bibinfo{year}{2002}).

\bibitem[{\citenamefont{{Particle Data Group, D.E.~Groom {\em et
  al.}}}(2000)}]{PDG}
\bibinfo{author}{\bibnamefont{{Particle Data Group, D.E.~Groom {\em et al.}}}},
  \bibinfo{journal}{Eur.~Phys.~J.~C} \textbf{\bibinfo{volume}{15}},
  \bibinfo{pages}{1} (\bibinfo{year}{2000}).

\bibitem[{\citenamefont{Mart and Bennhold}(2000)}]{Mart2}
\bibinfo{author}{\bibfnamefont{T.}~\bibnamefont{Mart}} \bibnamefont{and}
  \bibinfo{author}{\bibfnamefont{C.}~\bibnamefont{Bennhold}},
  \bibinfo{journal}{Phys.~Rev.~C} \textbf{\bibinfo{volume}{61}},
  \bibinfo{pages}{(R)012201} (\bibinfo{year}{2000}).

\bibitem[{\citenamefont{Janssen et~al.}(2001)\citenamefont{Janssen, Ryckebusch,
  {Van~Nespen}, Debruyne, and {Van~Cauteren}}}]{Janssen_role_hyp}
\bibinfo{author}{\bibfnamefont{S.}~\bibnamefont{Janssen}},
  \bibinfo{author}{\bibfnamefont{J.}~\bibnamefont{Ryckebusch}},
  \bibinfo{author}{\bibfnamefont{W.}~\bibnamefont{{Van~Nespen}}},
  \bibinfo{author}{\bibfnamefont{D.}~\bibnamefont{Debruyne}}, \bibnamefont{and}
  \bibinfo{author}{\bibfnamefont{T.}~\bibnamefont{{Van~Cauteren}}},
  \bibinfo{journal}{Eur.~Phys.~J.~A} \textbf{\bibinfo{volume}{11}},
  \bibinfo{pages}{105} (\bibinfo{year}{2001}).

\bibitem[{\citenamefont{Lomon}(2001)}]{Lomon}
\bibinfo{author}{\bibfnamefont{E.~L.} \bibnamefont{Lomon}},
  \bibinfo{journal}{Phys.~Rev.C} \textbf{\bibinfo{volume}{64}},
  \bibinfo{pages}{035204} (\bibinfo{year}{2001}).

\bibitem[{\citenamefont{M\"unz et~al.}(1995)\citenamefont{M\"unz, Resag,
  Metsch, and Petry}}]{Munz}
\bibinfo{author}{\bibfnamefont{C.}~\bibnamefont{M\"unz}},
  \bibinfo{author}{\bibfnamefont{J.}~\bibnamefont{Resag}},
  \bibinfo{author}{\bibfnamefont{B.}~\bibnamefont{Metsch}}, \bibnamefont{and}
  \bibinfo{author}{\bibfnamefont{H.}~\bibnamefont{Petry}},
  \bibinfo{journal}{Phys.~Rev.~C} \textbf{\bibinfo{volume}{52}},
  \bibinfo{pages}{2110} (\bibinfo{year}{1995}).

\bibitem[{\citenamefont{Merten et~al.}(2002)\citenamefont{Merten, L\"oring,
  Kretzschmar, Metsch, and Petry}}]{Merten}
\bibinfo{author}{\bibfnamefont{D.}~\bibnamefont{Merten}},
  \bibinfo{author}{\bibfnamefont{U.}~\bibnamefont{L\"oring}},
  \bibinfo{author}{\bibfnamefont{K.}~\bibnamefont{Kretzschmar}},
  \bibinfo{author}{\bibfnamefont{B.}~\bibnamefont{Metsch}}, \bibnamefont{and}
  \bibinfo{author}{\bibfnamefont{H.-R.} \bibnamefont{Petry}},
  \bibinfo{journal}{Eur. Phys. J. A} \textbf{\bibinfo{volume}{14}},
  \bibinfo{pages}{477} (\bibinfo{year}{2002}).

\bibitem[{\citenamefont{Gross and Riska}(1987)}]{Gross}
\bibinfo{author}{\bibfnamefont{F.}~\bibnamefont{Gross}} \bibnamefont{and}
  \bibinfo{author}{\bibfnamefont{D.}~\bibnamefont{Riska}},
  \bibinfo{journal}{Phys.~Rev.~C} \textbf{\bibinfo{volume}{36}},
  \bibinfo{pages}{1928} (\bibinfo{year}{1987}).

\bibitem[{\citenamefont{Feuster and Mosel}(1999)}]{Feuster2}
\bibinfo{author}{\bibfnamefont{T.}~\bibnamefont{Feuster}} \bibnamefont{and}
  \bibinfo{author}{\bibfnamefont{U.}~\bibnamefont{Mosel}},
  \bibinfo{journal}{Phys.~Rev.~C} \textbf{\bibinfo{volume}{59}},
  \bibinfo{pages}{460} (\bibinfo{year}{1999}).

\bibitem[{\citenamefont{Haberzettl}(1997)}]{Haberzettl_gauge}
\bibinfo{author}{\bibfnamefont{H.}~\bibnamefont{Haberzettl}},
  \bibinfo{journal}{Phys.~Rev.~C} \textbf{\bibinfo{volume}{56}},
  \bibinfo{pages}{2041} (\bibinfo{year}{1997}).

\bibitem[{\citenamefont{Davidson and Workman}(2001)}]{Davidson}
\bibinfo{author}{\bibfnamefont{R.}~\bibnamefont{Davidson}} \bibnamefont{and}
  \bibinfo{author}{\bibfnamefont{R.}~\bibnamefont{Workman}},
  \bibinfo{journal}{Phys.~Rev.~C} \textbf{\bibinfo{volume}{63}},
  \bibinfo{pages}{025210} (\bibinfo{year}{2001}).

\bibitem[{\citenamefont{Hsiao et~al.}(2000)\citenamefont{Hsiao, Lu, and
  Yang}}]{Hsiao}
\bibinfo{author}{\bibfnamefont{S.}~\bibnamefont{Hsiao}},
  \bibinfo{author}{\bibfnamefont{D.}~\bibnamefont{Lu}}, \bibnamefont{and}
  \bibinfo{author}{\bibfnamefont{S.}~\bibnamefont{Yang}},
  \bibinfo{journal}{Phys.~Rev.~C} \textbf{\bibinfo{volume}{61}},
  \bibinfo{pages}{068201} (\bibinfo{year}{2000}).

\bibitem[{\citenamefont{Haberzettl et~al.}(1998)\citenamefont{Haberzettl,
  Bennhold, Mart, and Feuster}}]{Haberzettl}
\bibinfo{author}{\bibfnamefont{H.}~\bibnamefont{Haberzettl}},
  \bibinfo{author}{\bibfnamefont{C.}~\bibnamefont{Bennhold}},
  \bibinfo{author}{\bibfnamefont{T.}~\bibnamefont{Mart}}, \bibnamefont{and}
  \bibinfo{author}{\bibfnamefont{T.}~\bibnamefont{Feuster}},
  \bibinfo{journal}{Phys.~Rev.~C} \textbf{\bibinfo{volume}{58}},
  \bibinfo{pages}{(R)40} (\bibinfo{year}{1998}).

\bibitem[{\citenamefont{David et~al.}(1996)\citenamefont{David, Fayard, Lamot,
  and Saghai}}]{David}
\bibinfo{author}{\bibfnamefont{J.}~\bibnamefont{David}},
  \bibinfo{author}{\bibfnamefont{C.}~\bibnamefont{Fayard}},
  \bibinfo{author}{\bibfnamefont{G.}~\bibnamefont{Lamot}}, \bibnamefont{and}
  \bibinfo{author}{\bibfnamefont{B.}~\bibnamefont{Saghai}},
  \bibinfo{journal}{Phys.~Rev.~C} \textbf{\bibinfo{volume}{53}},
  \bibinfo{pages}{2613} (\bibinfo{year}{1996}).

\end{thebibliography}

\end{document}